\documentclass[twocolumn,amsmath,showkeys,showpacs,prl]{revtex4-1}
\usepackage{dcolumn}% Align table columns on decimal point
\usepackage{bm}% bold math
\usepackage[T1]{fontenc}
\usepackage{amsmath}
\usepackage{graphicx}
\usepackage{tabularx}

\usepackage{hyperref}
\hypersetup{colorlinks=true, linkcolor=blue, citecolor=blue, urlcolor=blue, pdftitle={Strain engineering multiferroism in \emph{Pnma} NaMnF$_3$ fluoroperovskite}, pdfauthor={A. C. Garcia-Castro}}

\begin{document}
\title{Strain engineering multiferroism in \emph{Pnma} NaMnF$_3$ fluoroperovskite}
\author{A. C. Garcia-Castro$^{1,2}$}
\email{a.c.garcia.castro@gmail.com}
\author{A. H. Romero$^{3,2}$}
\email{alromero@mail.wvu.edu}
\author{E. Bousquet$^{1}$}
\email{eric.bousquet@ulg.ac.be}
\affiliation{$^1$Physique Th\'eorique des Mat\'eriaux, Universit\'e de Li\`ege, B-4000 Sart-Tilman, Belgium}
\affiliation{$^2$Centro de Investigaci\'on y Estudios Avanzados del IPN, MX-76230, Quer\'etaro, M\'exico}
\affiliation{$^3$Physics Department, West Virginia University, WV-26506-6315, Morgantown, USA}

\begin{abstract}
In this study we show from first principles calculations the possibility to induce multiferroic and magnetoelectric functional properties in the \emph{Pnma} NaMnF$_3$ fluoroperovskite by means of epitaxial strain engineering.
Surprisingly, we found a very strong non-linear polarization-strain coupling that drives an atypical amplification of the ferroelectric polarization for either compression or expansion of the cell. 
This property is associated with a non-collinear antiferromagnetic ordering, which induces a weak ferromagnetism and thus, making the strained NaMnF$_3$ fluoroperovskite multiferroic.
We also calculate the magnetoelectric response and we found it to be composed by linear and non-linear components with amplitudes similar to the ones of Cr$_2$O$_3$.
These properties show that it is possible to move the fluoride family toward functional applications with unique responses.
\end{abstract}

\maketitle

%\section{Introduction}
The search for new and innovative materials with promising multifunctional multiferroic (MF) properties has been one of the keystones of condensed matter research in the last decade \cite{Eerenstein2006a, martin2010b}. 
Several systems and compounds based on oxide perovskites have been reported as ideal candidates. 
In these systems, even in the absence of a ferroic order at the bulk level, physical constrains such as biaxial epitaxial strain in thin films can be used to artificially induce a ferroelectric (FE) order  \cite{Rabe2005}. 
This has been successfully performed in paraelectric crystals such as SrTiO$_3$ \cite{Haeni2004} or CaTiO$_3$ \cite{Eklund2009}.
A similar mechanism has been reported for magnetic perovskites \cite{Spaldin2005strain, Gunter2012, Bousquet2011}, and thus exceeding the so-called d$^0$-ness rule that is expected to prevent the formation of a FE phase in magnetic perovskites \cite{Hill2000a}.  
 However, the possibility for new materials with MF properties in new stoichiometries away from the easily polarizable oxides is still evasive. 
Scott and Blinc have reported several other possible and unexplored MF and magnetoelectric (ME) candidates in fluoride crystal class of materials \cite{Scott2011}. 
Nonetheless, none of the reported fluoride candidates belongs to the most claimed perovskite family.
Recently, we have shown that even if none of the fluoroperovskites is reported with a FE ground state (except CsPbF$_3$ \cite{0953-8984-13-22-305, Smith2015})  they have nevertheless the propensity to have a FE instability in their high symmetry cubic reference structure \cite{acgarciacastro2014}. 
We have identified that the FE instability of the fluoroperovskites is related to a steric geometric effect when small cations lie at the $A$-site in opposition to the charge transfer origin observed in the oxides.
Unfortunately, these fluoroperovskites keep the ``undesired'' competition between the FE and antiferrodistortive (AFD) instabilities such as the AFD dominate in the bulk ground states.
Interestingly, and similarly to the oxides, we have shown that playing with epitaxial strain can also induce FE orders in fluoroperovskites such as it opens the way of discovering new ferroelectrics of geometric origin in the ionic crystals. 
In this paper, we explore from first-principles the properties of strain-induced ferroelectricity and multiferroism in NaMnF$_3$ and find breakthrough differences with the oxides. 
Interestingly, we show that the geometric origin of the ferroelectricity drives unique responses such as a non-linear strain-polarization coupling that is associated with a very strong second order piezoelectric response and a non-linear ME effect.
We also predict a very strong spin-canting in the antiferromagnetic NaMnF$_3$ crystal that drives a sizeable ferromagnetic component and thus making strained NaMnF$_3$ a good MF candidate.
All of that shows that engineering ferroelectricity and multiferroism in the fluoroperovskite class of materials is very appealing to discover unexplored multifunctional properties with unprecedented responses.

%\section{Computational Details} 
We used Density Functional Theory (DFT) within the Projector Augmented Wave (PAW) \cite{Blochl1994} method as implemented in the Vienna Ab-initio Simulation Package (VASP) \cite{Kresse1996,Kresse1999} . 
Seven valence electrons for Na (2p$^6$3s$^1$), thirteen for Mn (3p$^6$4s$^2$3d$^5$), and seven for F (2s$^2$2p$^5$) were taken into account in the pseudo-potential. 
The exchange correlation was represented within the General Gradient Approximation (GGA) by Perdew-Burke-Ernkzerhof for Solids (PBEsol) parameterization \cite{Perdew2008} and corrected with the DFT$+U$  method \cite{Liechtenstein1995} ($U$ = 4.0 eV) in order to treat the localised $d$ electrons of Mn. 
The periodic solution of these crystalline structures was represented by using Bloch states with a Monkhorst-Pack \emph{k}-point mesh of 6$\times$4$\times$6 and 700 eV energy cut-off, which give forces converged to less than 1 meV$\cdot$\r{A}$^{-1}$.  
The spin-orbit coupling (SOC) was included to simulate the non-collinear calculations \cite{Hobbs2000}. 
Born effective charges and phonon calculations were performed with the density functional perturbation theory (DFPT) \cite{gonze1997} as implemented in VASP.  
The FE spontaneous polarization was computed through the Berry phase approach \cite{Vanderbilt2000}. 
The ME coupling was obtained by computing the spontaneous polarization as a function of the applied Zeeman magnetic field as implemented by \citeauthor{PRL-delaney2011} \cite{PRL-delaney2011} within the LDA approach.

%%%%%%%%%%%%%%%%

%\section{Results and Discussion}
Sodium-manganese fluoride (NaMnF$_3$) crystallizes in the \emph{Pnma} (group No. 62) structure at room conditions. 
Within our DFT calculations we obtained relaxed cell parameters $a_0$ = 5.750  \r{A}, $b_0$ = 8.007  \r{A}  and $c_0$ = 5.547 \r{A} in good agreement with experimental reports by Daniel \emph{et. al.} \cite{Daniel1995a} with a maximum error of 0.2\%.
This non-polar ground state is coming from the condensation of AFD modes and anti-polar displacements of the Na atoms such as the FE instability observed in the cubic phase \cite{acgarciacastro2014} is suppressed by the AFD ones.
Thus, the competition between FE, anti-polar displacements and AFD distortions monitors the \emph{Pnma} phase but the balance between them is delicate in the case of NaMnF$_3$ since we still find a very soft $B_{2u}$ polar mode at 18 cm$^{-1}$. 
This means that the \emph{Pnma} phase is very close to be FE and thus close to be an incipient ferroelectric \cite{Rabe2007book}. 
This property can be verified experimentally.

%%%%%%%%%%%%%%%%

\emph{Ferroelectric Behavior in Strained NaMnF$_3$}.--- 
In FE oxides it is well known that the epitaxial strain can induce ferroelectricity in the \emph{Pnma} phase \cite{Bousquet2011,Eklund2009} 
or enhance it in FE compounds such as PbTiO$_3$ or BiFeO$_3$ \cite{Spaldin2005strain,Rabe2005}.  
Here we show that a similar strain engineering ferroelectricity can be used in NaMnF$_3$. 
We suppose a cubic perovskite substrate ($a_c$) as the source of the strain by imposing to the $Pnma$ crystal $a$ = $c$ = $\sqrt 2a_c$. 
We choose the 0\% strain reference to be $a_r=(a_0+c_0)/2$ with $a_0$ and $c_0$ the unstrained relaxed cell parameters of the \emph{Pnma} phase and defining the strain amplitude by $\varepsilon = (a-a_r)/a_r$.  
In this configuration the soft $B_{2u}$ mode is polarized perpendicular to the biaxial strain (\emph{i.e.} along the orthorhombic $b$-axis).

 \begin{figure}[htb]
 \centering
 \includegraphics[width=8.6cm,keepaspectratio=true]{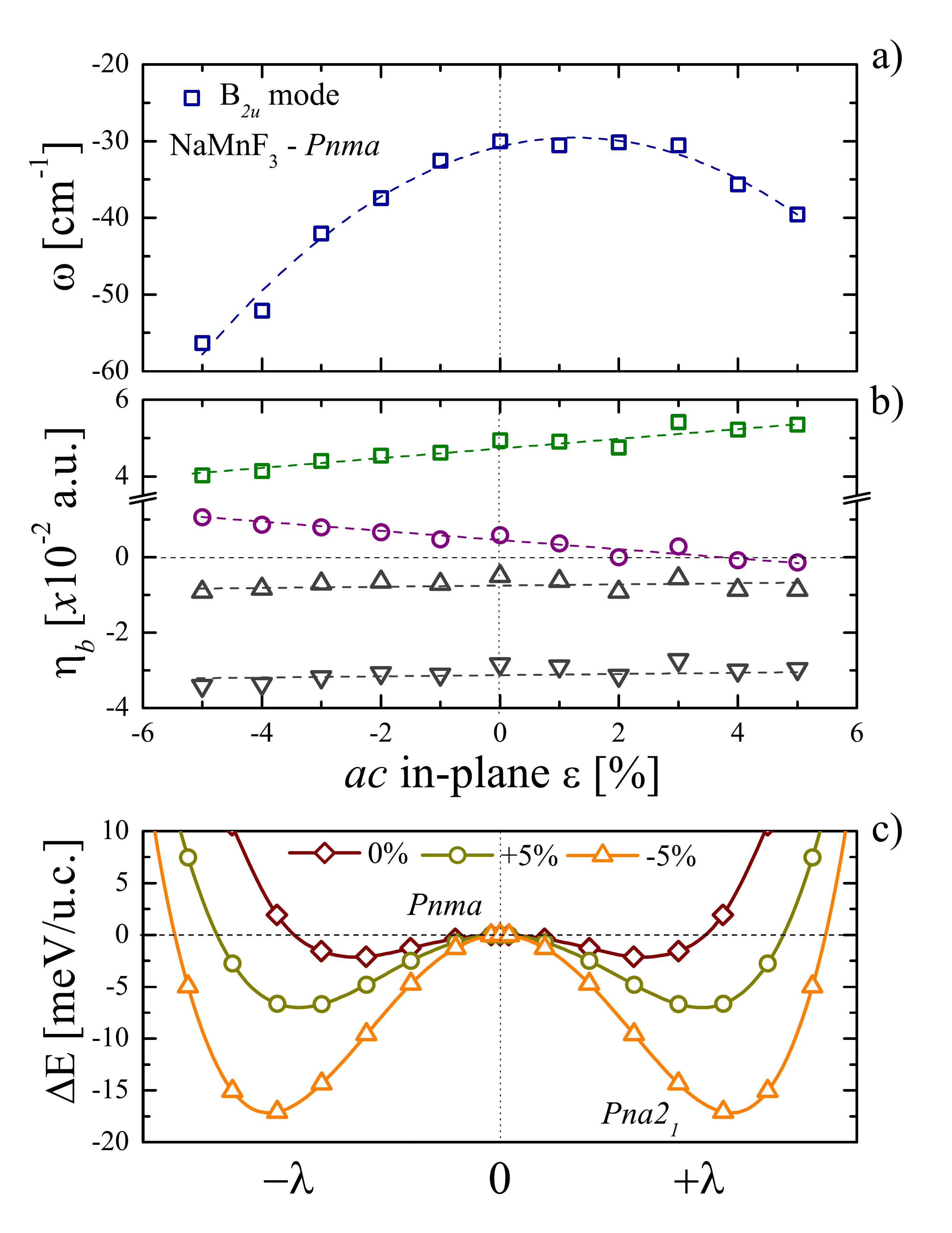}
 \caption{(Color online) a) Frequency of the polar $B_{2u}$ unstable mode of \emph{Pnma} NaMnF$_3$ with respect to the epitaxial strain (imaginary frequencies are represented as negative values). 
 b) $B_{2u}$ mode eigendisplacements contributions of  each atomic type (Na in green squares, Mn in purple circles and  F$_\perp$, F$_\parallel$  in upper and down gray triangles respectively). 
 c) Energy versus freezing-in amplitude of the $B_{2u}$ mode for 0\%, +5\% and -5\% strain values, which shows that expansion or compression enlarge ferroelectricity.}
 \label{fig:FE-vs-n}
\end{figure}

In Figure \ref{fig:FE-vs-n}a we plot the evolution of the $B_{2u}$ mode frequency with respect to the epitaxial strain.
Unexpectedly, we see that the $B_{2u}$ mode becomes unstable \emph{whatever} the value of the epitaxial strain, in compression or in expansion.
This means that for any value of the epitaxial train, NaMnF$_3$ has a FE instability.
When condensing different amplitudes of this mode we have the double well plotted in \ref{fig:FE-vs-n}c, which shows that for either positive or negative strain amplitudes, the FE double well is amplified.
When performing the structural relaxations, we indeed find a FE ground state with the $Pna2_1$ space group (No. 33) for all strain values. 
Then, the ground state of epitaxially constrained NaMnF$_3$ is always FE and thus MF due to the magnetically active  Mn$^{+2}$ cation. 
The strain-induced ferroelectricity in \emph{Pnma} perovskites is well established in oxides \cite{Eklund2009,Haeni2004,Bousquet2011} and can thus be extended to the fluoride family but with the following striking differences.
The first one is that the polarization is enhanced for positive or negative values of the strain while in oxide the relationship is linear such as if one direction of strain enhance the FE polarization, it reduces and destroys it for the opposite direction \cite{Rabe2005}.
A second difference is that the polarization develops in the direction where the anti-polar motions of the Na atoms ($X_5^+$ mode) are absent. 
This property is related to the geometric origin of the polar instability in fluoroperovskites.
The Na is strongly unstable due to the loss of bonding of this small cation in a large unit cell.
This loss of bonding is relaxed by the motion of the Na away from its high symmetry position through polar or anti-polar motions.
In the \emph{Pnma} phase this is done through anti-polar motions along the $a$ and $c$ directions but along the $b$ direction the loss of bonding is still present and it is the source of the softness of the polar $B_{2u}$ mode along the $b$ direction.

To understand the origin of these unusual FE responses, we analysed the eigendisplacements ($\eta_b$) changes of the unstable mode $B_{2u}$ under epitaxial strain (see Fig. \ref{fig:FE-vs-n}b). 
We can see that the Na contribution is reduced when going from positive to negative values of the strain while it is the opposite for the Mn atom, the fluorine contributions being unchanged in the same range. 
This means that the change of mode frequency is related to the change in the eigendisplacement pattern from Na dominated to Na+Mn dominated.

 \begin{figure}[htb]
 \centering
 \includegraphics[width=8.8cm,keepaspectratio=true]{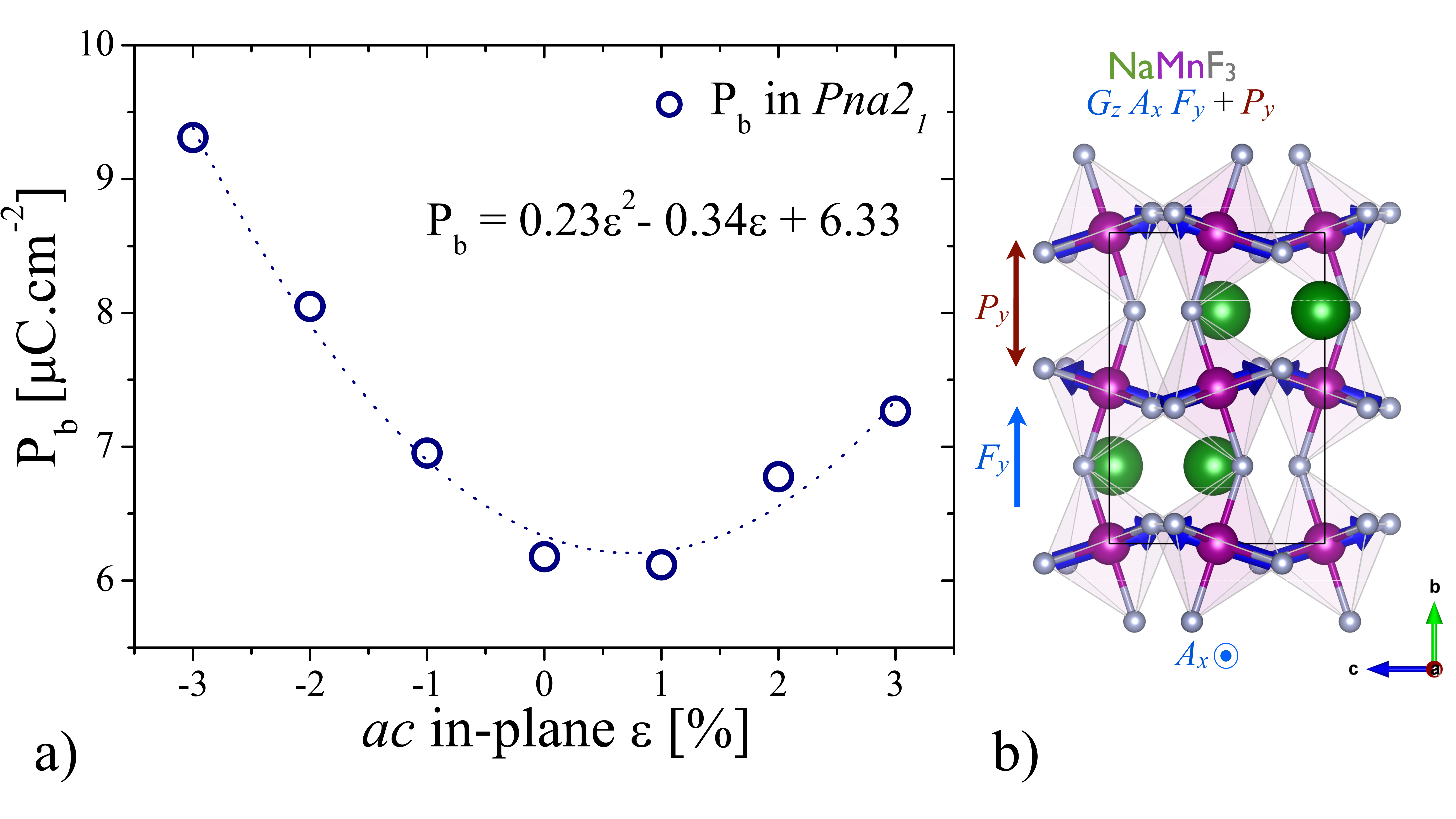}
 \caption{(Color online) a) Evolution of the polarization along $b$-axis of $Pna2_1$ NaMnF$_3$ with respect to the epitaxial strain.
  A unusual non-linear polarization strain  coupling is observed and can be associated to a non-linear piezoelectric response. 
  b) Schematic view of the $Pna2_1$ NaMnF$_3$  structure in which Na, Mn and F ions are depicted in green, violet and grey respectively. 
  The non-collinear magnetic ground state  (large arrows going through the Mn), direction of the cantings as well as the electric polarization along the $b$-axis are also pictured.}
 \label{fig:pol-vs-n}
\end{figure}

%%%%%%%%%%%%%%%%%%%%%%%

In Fig. \ref{fig:pol-vs-n} we report the evolution of the polarization versus the epitaxial strain.
Here again, we observe an unusual and outstanding polarization/strain coupling in which the non-linear contribution dominates the linear one.
This can be expressed in terms of the strain, piezoelectric constants and polarization as follows:
\begin{equation}
\label{eqn:eq1}% requires amsmath; align* for no eq. number
   P_{\mu} = \sum_{\substack{j\\}} e_{\mu j}\varepsilon_{j}  + \frac{1}{2} \sum_{\substack{j,k\\}} B_{\mu j k}\varepsilon_{j} \varepsilon_{k} 
\end{equation}
where P$_{\mu}$ is the spontaneous polarization (with $\mu$ = 1, 2, 3 for the cartesian components \emph{x, y, z} respectively), $\varepsilon_j$ ($j$ = 1, 2,...., 6 as the components of the tensor into the Voigt notation) is the strain tensor and $e_{\mu j}$, $B_{\mu j k}$ are the linear and quadratic piezoelectric coefficients respectively. 
The computed linear piezoelectric coefficients $e_{21}$, $e_{23}$, $e_{22}$, $e_{16}$, and $e_{34}$ at 0\% strain are equal to 1.010, -0.672, 0.064, -0.108, and 0.058 C$\cdot$m$^{-2}$ respectively and are not far from the ones observed in BaTiO$_3$ \cite{PhysRevB.83.054112,Furuta20102350}. From the quadratic fitting observed in Fig \ref{fig:pol-vs-n}, we can extract that the overall non-linear contribution related to the sum of the $B_{\mu j k}$ components and we found it to be about 60\% of the linear piezoelectric constants ($e_{\mu j}$).
This unique large non-linear piezoelectric response can thus impact novel applications of this material.
Such a non-linear piezoelectric property has been found only in a very few compounds, such as some zinc-blende semiconductors \cite{Bester2006}.

%%%%%%%%%%%%%%%%%%%%

\emph{Non-collinear magnetism and ME coupling in  $Pna2_1$ NaMnF$_3$}.--- 
In the following section, we analyse the magnetic properties of the strained NaMnF$_3$. 
The possible allowed magnetic orderings and couplings were obtained based on group theory analysis \cite{bertaut}, 
which we report in Table \ref{tab:ord-sym}.
We note that in the $Pna2_1$ phase, whatever the magnetic states adopted by the system, spin canting and ME responses are allowed by symmetry.
Experimentally, it has been observed in the \emph{Pnma} phase a predominant $G$-type AFM behavior with spin canting driving a weak magnetization along $b$-axis \cite{Shane1967a}.
Our non-collinear calculations give the same non-collinear magnetic ground state with a marked $G$-type AFM along the $c$-axis, $A$-type AFM canting along the $a$ direction and a ferromagnetic canting along the $b$ direction (\emph{A$_x$F$_y$G$_z$} in Bertaut's notation). 
In the strained $Pna2_1$ phase we do not observe a magnetic transition, then, the system always keeps its \emph{G$_z$A$_x$F$_y$} non-collinear ground state.
The ferromagnetic canting gives a magnetization of 0.02 $\mu_B$/atom, which is one order of magnitude larger than the one found in CaMnO$_3$ (0.004 $\mu_B$/atom \cite{Bousquet2011}).

The related magnetic point group (\emph{m'm'2}) of the \emph{A$_x$F$_y$G$_z$} magnetic ground state, allows for linear and non-linear ME response \cite{magnetic-tables-2006} such as the energy expansion can be written as follows:
\begin{equation}
\label{eqn:eq3}
P_i=\alpha_{ik}H_k + \frac{1}{2} \beta_{ijk}H_jH_k + \gamma_{jik}H_jE_k
\end{equation}
where $P_i$ is the spontaneous ferroelectric polarization along each particular cartesian axis ($i$ = $x$, $y$ and $z$), $H_k$ is the applied magnetic field along the $k$ axis and $\alpha_{ik}$ and $\beta_{ijk}$/$\gamma_{jik}$ are the linear and non-linear ME tensor components respectively.
 
\begin{table}[htbp!]
\caption{Allowed magnetic orderings and ME response in the \emph{D$_{2h}$} point symmetry group \cite{Arroyo1, Arroyo2} of the $Pna2_1$ phase. }
\centering
\begin{tabularx}{\columnwidth}{c  c  c c}
\hline
\hline
Magnetic      &   \multicolumn{3}{c}{$Pna2_1$} \rule[-1ex]{0pt}{3.5ex} \\
Ordering & Character &  linear ME  & second order ME \rule[-1ex]{0pt}{3.5ex} \\
\hline
\emph{C$_x$, G$_y$, F$_z$}   &\emph{A$_2$} & $\alpha_{yz}$, $\alpha_{zy}$ & $\beta_{ijk}$, $\gamma_{jik}$ \rule[-1ex]{0pt}{3.5ex}\\
\emph{A$_x$, F$_y$, G$_z$}   &\emph{B$_1$}  & $\alpha_{xx}$, $\alpha_{yy}$, $\alpha_{zz}$ & $\beta_{ijk}$, $\gamma_{jik}$ \rule[-1ex]{0pt}{3.5ex}\\
\emph{F$_x$, A$_y$, C$_z$}   &\emph{B$_2$}  & $\alpha_{xz}$, $\alpha_{zx}$ & $\beta_{ijk}$, $\gamma_{jik}$  \rule[-1ex]{0pt}{3.5ex}\\
\emph{G$_x$, C$_y$, A$_z$}  &\emph{A$_1$}  & $\alpha_{xy}$, $\alpha_{yx}$  &  $\beta_{ijk}$, $\gamma_{jik}$ \rule[-1ex]{0pt}{3.5ex}\\
\hline
\hline
\end{tabularx}
\label{tab:ord-sym}
\end{table}

\begin{figure}[htb]
 \centering
 \includegraphics[width=8.5cm,keepaspectratio=true]{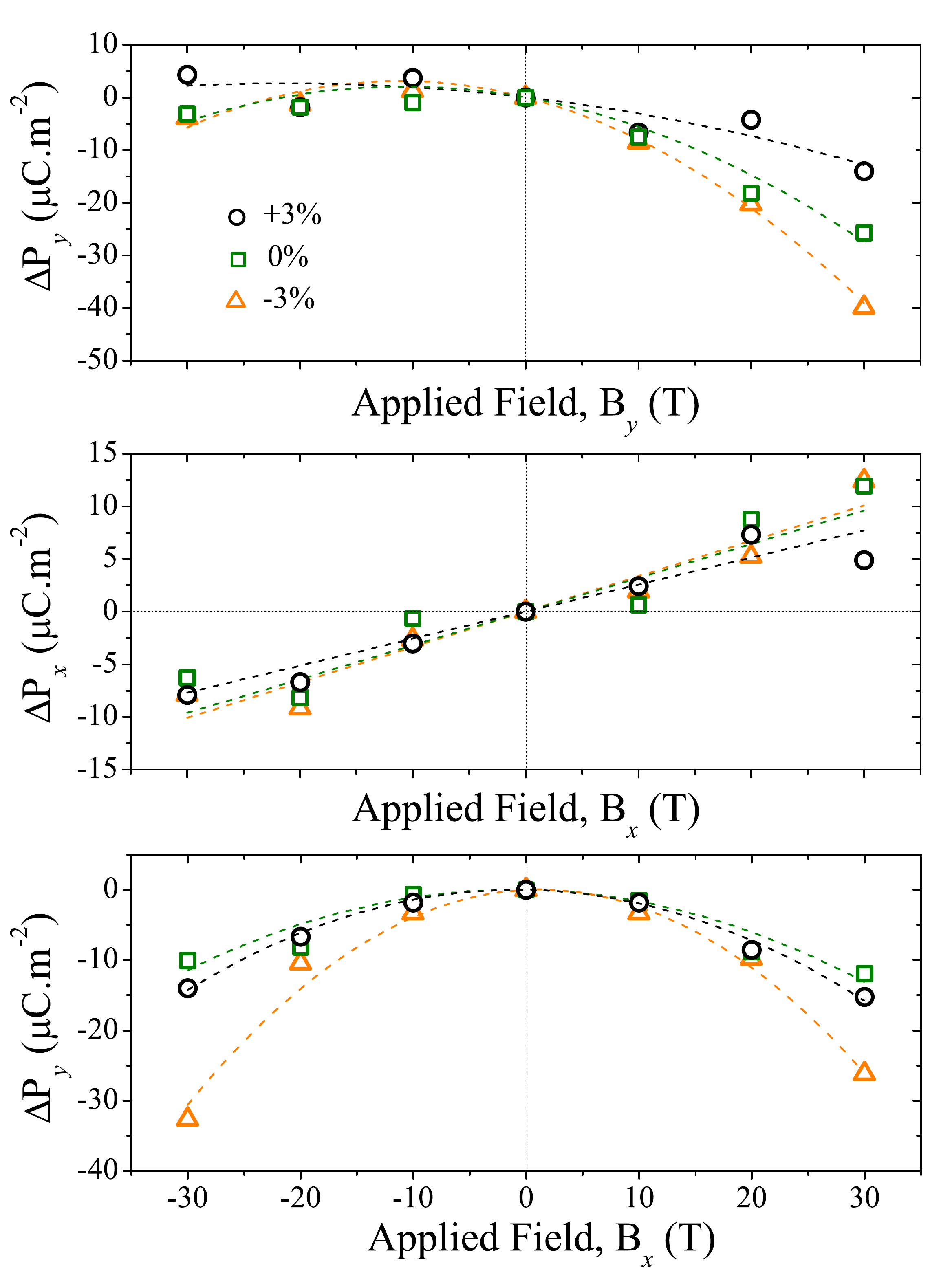}
 \caption{(Color online) Change of polarization ($\Delta P_i$) versus magnetic field in strained NaMnF$_3$  ($Pna2_1$ phase). 
 The dashed lines represent fitting curves following Eq. \ref{eqn:eq3}. 
 The second-order ME coupling can be appreciated from the figure for fields applied along $x$ and $y$ directions an increase of the non-linear behavior is observed when strain goes from positive to negative values. 
 The first and second order ME tensor components computed from these plots are presented in the Table \ref{tab:ME-coupling}.}
 \label{fig:ME}
\end{figure}

In Figure \ref{fig:ME} we report the evolution of the electric polarization with respect to the amplitude of the applied magnetic field along different directions. In Table \ref{tab:ME-coupling} we report the extracted ME coefficients at three different epitaxial strains.
Our results reveal that a sizeable non-linear ME coupling is present when the magnetic field is applied along the $y$-axis (parallel to the weak-FM moment).
When the field is applied along the $x$-axis, we observe a linear ME response along the same direction ($\alpha_{xx}$) and a non-linear one along the $y$-axis ($\beta_{yxx}$).
The amplitude of the linear ME response of strained NaMnF$_3$ is close to the one reported in Cr$_2$O$_3$ \cite{PhysRevLett.6.607, Cr2O3-1979, Cr2O3-1999, PhysRevLett.101.117201}. 
All the ME coefficients increase when the strain goes from positive to negative values, which is consistent with the fact that the polarization contribution of the magnetic Mn atoms is larger for compressive strains and thus we can expect a stronger coupling between electric polarization and magnetism.
Additionally, the global forms of the ME responses in this fluoroperovskite is very similar to the one predicted for CaMnO$_3$  \cite{Bousquet2011}, which shows that this might be a general rule for strain-induced FE in the $Pnma$ structure.

\begin{table}[htbp!]
\caption{ME coefficients of strained NaMnF$_3$ expressed as in the Eq. \ref{eqn:eq3}, $\alpha_{ik}$ in [ps$\cdot$m$^{-1}$] and  $\beta_{jik}$ in  [$\times$10$^{-8}$ ps$\cdot$A$^{-1}$]. 
}
\centering
\begin{tabularx}{\columnwidth}{X  X  X | X  X  X}
\hline
\hline
$\varepsilon$ [\%] &  $\alpha_{yy}$ & $\alpha_{xx}$ & $\beta_{yyy}$  &  $\beta_{xxx}$  & $\beta_{yxx}$  \rule[-1ex]{0pt}{3.5ex} \\
\hline
+3$\%$ & -0.314  & 0.322  &  -0.927  & 0.000 & -2.625  \rule[-1ex]{0pt}{3.5ex}\\
0$\%  $ & -0.417  &  0.403  &  -2.923  & 0.000 & -2.156   \rule[-1ex]{0pt}{3.5ex}\\
--3$\%$ & -0.698  &  0.423 & -3.931   & 0.000 & -4.973  \rule[-1ex]{0pt}{3.5ex}\\
\hline
\hline
\end{tabularx}
\label{tab:ME-coupling}
\end{table}

%%%%%%%%%%%%%%%%%%%%

%\section{Conclusions}
From the industrial point of view, fluorides have proven to be of high interest for numerous long term applications, such as fluorides-based glasses with a large thermal expansion, low refractive and non-linear index \cite{VIDEAU1985309,RAVEZ1985469}, strong magnets with an optically transparency in the visible light \cite{wolfe1970,DANCE1985371}, electrochemical devices, solid-state batteries, gas sensors, and electrochromic systems \cite{REAU1985423} or catalyst surfaces in base of metal fluorides such as AlF$_3$ \cite{func-fluorides-cap3}. 
Thus, adding multiferroic properties in fluoroperovskites would open an exciting opportunity to their use in novel and extended applications.

In this letter we have shown from first principles calculations that a multiferroic ground state can be induced in NaMnF$_3$ thin films through epitaxial strain with an unprecedented non-linear polarization/strain coupling.
This unusual polarization/strain coupling is related to the specific geometric source of ferroelectricity of this ionic crystal and further analysis should be performed on other similar systems in order to confirm the rule (for example in the Ba$M$F$_4$ family).
We have shown that this unique multiferroic state drives large and uncommon second order piezoelectric response and a non-linear magnetoelectric response.
We hope our results will motivate further theoretical and experimental studies owing to the fact that making a fluoroperovkite multiferroic is a primer and it would open the field of multifunctional materials to new candidates with novel and remarkable responses.

%%%%%%%%%%%%%%%%%%%%
\begin{acknowledgments}
\emph{Acknowledgements:}
This work used the Extreme Science and Engineering Discovery Environment (XSEDE), which is supported by National Science Foundation grant number OCI-1053575. In Belgium, the computational resources have been provided by the Consortium des Equipements de Calcul Intensif (CECI), funded by the F.R.S.-FNRS under Grant No. 2.5020.11. Additionally, the authors acknowledge the support from the and Texas Advances Computer Center (TACC). This work was supported by FRS-FNRS Belgium (EB). AHR acknowledge the support of DMREF-NSF 1434897 and the Donors of the American Chemical Society Petroleum Research Fund for partial support of this research under contract 54075-ND10.
\end{acknowledgments}

\bibliography{library}

\end{document}